%% file: 143.tex
\newif\ifdraft\draftfalse
\documentclass{llncs}

\input{prelude}

\title{Synthesizing Probabilistic Invariants \\ via Doob's Decomposition}
\titlerunning{Synthesizing Probabilistic Invariants via Doob's Decomposition}
\author{Gilles Barthe\inst{1}
  \and Thomas Espitau\inst{2}
  \and Luis Mar{\'i}a Ferrer{ }Fioriti\inst{3}
  \and Justin Hsu\inst{4}}

\authorrunning{G. Barthe, T. Espitau, L. M. Ferrer{ }Fioriti, J. Hsu} % abbreviated author list (for running head)
\institute{
  $\mbox{}^1$ IMDEA Software Institute \quad
  $\mbox{}^2$ ENS Cachan \\
  $\mbox{}^3$ Saarland University \quad
  $\mbox{}^4$ University of Pennsylvania
}

\begin{document} 
\maketitle

\begin{abstract}
  When analyzing probabilistic computations, a powerful approach is to first find
  a \emph{martingale}---an expression on the program variables
  whose expectation remains invariant---and then apply the optional stopping
  theorem in order to infer properties at termination time.
  One of the main challenges, then, is to systematically find martingales.

  We propose a novel procedure to synthesize martingale expressions from an
  arbitrary initial expression. Contrary to state-of-the-art approaches, we do
  not rely on constraint solving. Instead, we use a symbolic construction based
  on \emph{Doob's decomposition}. This procedure can produce very complex
  martingales, expressed in terms of conditional expectations.

  We show how to \emph{automatically} generate and simplify these martingales,
  as well as how to apply the \emph{optional stopping theorem} to infer properties at
  termination time. This last step typically involves some simplification steps,
  and is usually done manually in current approaches. We implement our
  techniques in a prototype tool and demonstrate our process on several
  classical examples. Some of them go beyond the capability of current
  semi-automatic approaches.
\end{abstract}

\section{Introduction}

Probabilistic computations are a key tool in modern computer science.
They are ubiquitous in machine learning, privacy-preserving data
mining, cryptography, and many other fields. They are also a common
and flexible tool to model a broad range of complex real-world
systems. Not surprisingly, probabilistic computations have been
extensively studied from a formal verification perspective. However,
their verification is particularly challenging.

In order to understand the difficulty, consider the standard way to
infer properties about the final state of a non-probabilistic program
using a strong invariant (an assertion which is preserved throughout
program execution) and a proof of termination. This proof principle is
not easily adapted to the probabilistic case. First, probabilistic
programs are interpreted as distribution transformers~\citep{Kozen81}
rather than state transformers. Accordingly, assertions (including
strong invariants) must be interpreted over distributions. Second, the
notion of termination is different for probabilistic programs.  We are
usually not interested in proving that \emph{all} executions are
finite, but merely that the probability of termination is $1$, a
slightly weaker notion. Under this notion, there may be no finite
bound on the number of steps over all possible executions.  So, we
cannot use induction to transfer local properties to the end of the
program---more complex limiting arguments are needed.

We can avoid some of these obstacles by looking at the \emph{average}
behavior of a program. That is, we can analyze numerical expressions (over
program variables) whose average value is preserved.  These
expressions are known as martingales, and have several technical
advantages.  First, martingales are easy to manipulate symbolically
and can be checked locally. Second, the average value of martingale is
preserved at termination, even if the control-flow of the program is
probabilistic.  This fact follows from the \emph{optional stopping
  theorem} (OST), a powerful result in martingale theory.

While martingales are quite useful, they can be quite
non-obvious. Accordingly, recent investigation has turned to
automatically synthesizing martingales.  State-of-the-art frameworks
are based on constraint solving, and require the user to provide
either a template expression \citep{KatoenMMM10,ChakarovS13} or a
limit on the search space \citep{ChenHWZ15,ChakarovS14}. The main
advantage of such approaches is that they are generally
complete---they find \emph{all} possible martingales in the search
space. However, they have their drawbacks: a slightly wrong template
can produce no invariant at all, and a lot of search space may be
needed to arrive at the martingale.

We propose a framework that \emph{complements} current
approaches---we rely on purely symbolic methods instead of solving
constraints or searching.  We require the user to provide a ``seed''
expression, from which we \emph{always} generate a martingale.
Our approach uses \emph{Doob's decomposition theorem}, which gives a
symbolic method to construct a martingale from any sequence of random values.
Once we have the martingale, we can apply optional stopping to reason about
the martingale at loop termination.  While the martingale
and final fact may be quite complex, we can use non-probabilistic invariants
and symbolic manipulations to automatically simplify them.

We demonstrate our techniques in a prototype implementation,
implementing Doob's decomposition and the Optional Stopping
Theorem. Although these proof principles have been long known to
probability theory, we are the first to incorporate them into an
automated program analysis. Given basic invariants and hints, our
prototype generates martingales and facts for a selection of examples.

\section{Mathematical preliminaries}

We briefly introduce some definitions from probability theory required for our
technical development. We lack the space to discuss the definitions in-depth,
but we will explain informally what the various concepts mean in our setting.
Interested readers can find a more detailed presentation in any standard
probability theory textbook (e.g., \citet{Williams91}).

First, we will need some basic concepts from probability theory.

\begin{defi}
  Let $\Omega$ be the set of outcomes.
  \begin{itemize}
    \item A \emph{sigma algebra} is a set $\SigmaAlgebra$ of subsets of $\Omega$,
      closed under complements and countable unions, and countable intersections.
    \item A \emph{probability measure} is a countably additive mapping \(\Prob:
        \SigmaAlgebra \to [0, 1]\) such that $\Prob(\Omega) = 1$.
    \item A \emph{probability} space is a triple  $(\Omega, \SigmaAlgebra, \Prob)$.
  \end{itemize}
\end{defi}

Next, we can formally define stochastic processes. These constructions are
technical but completely standard.
\begin{defi}
  Let $(\Omega, \SigmaAlgebra, \Prob)$ be a probability space.
  \begin{itemize}
    \item A \emph{(real) random variable} is a function $X : \Omega \to
      \Reals$. $X$ is $\SigmaAlgebra$-\emph{measurable} if $X^{-1}((a, b]) \in
      \SigmaAlgebra$ for every $a, b \in \Reals$.
    \item A \emph{filtration} is a sequence $\{ \mathcal{F}_i
      \}_{i \in \Nats}$ of sigma algebras such that $\mathcal{F}_i \subseteq
      \mathcal{F}$ and $\mathcal{F}_{i - 1} \subseteq \mathcal{F}_i$ for every
      $i > 0$. When there is a fixed filtration on $\mathcal{F}$, we will often
      abuse notation and write $\mathcal{F}$ for the filtration.
    \item A \emph{stochastic process} adapted to filtration $\mathcal{F}$ is
      a sequence of random variables $\{ X_i \}_{i \in \Nats}$ such that
      each $X_i$ is $\mathcal{F}_i$-measurable.
  \end{itemize}
\end{defi}

Intuitively, we can think of $\Omega$ as a set where each element represents a
possible outcome of the samples. In our setting, grouping
samples according to the loop iteration gives a
natural choice for the filtration: we can take $\mathcal{F}_i$ to be the set of
events that are defined by samples in iteration $i$ or before. A stochastic
process $X$ is adapted to this filtration if $X_i$ is defined in terms of
samples from iteration $i$ or before. Sampled variables at step $i$ are
independent of previous steps, so they are not $\mathcal{F}_{i - 1}$-measurable.

\paragraph*{Expectation.}
To define martingales, we need to introduce expected values and conditional
expectations.  The \emph{expected value} of a random variable is defined as
\[
   \Exp{X} \triangleq \int_\Omega X \cdot \dif \Prob
\]
where \(\int\) is the Lebesgue integral~\citep{Williams91}.
We say that a random variable is \emph{integrable} if \(\Exp{|X|}\) 
is finite.
Given a integrable random variable \(X\) and a sigma algebra
\(\SigmaAlgebra[G]\), a \emph{conditional expectation} of \(X\) with respect to
\(\SigmaAlgebra[G]\) is a random variable \(Y\) such that \(Y\) is
\(\SigmaAlgebra[G]\)-measurable, and \(\Exp{X \cdot \indFn{A}} = \Exp{Y \cdot
    \indFn{A}}\) for all events \(A \in \SigmaAlgebra[G]\). (Recall that the
\emph{indicator function} $\indFn{A}$ of an event $A$ maps $\omega \in A$ to
$1$, and all other elements to $0$.) Since one can show that this $Y$ is
essentially unique, we denote it by \(\CondExp{X}{\SigmaAlgebra[G]}\).

\paragraph*{Moments.} Our method relies on computing higher-order moments.
Suppose $X$ is a random variable with distribution $d$.  If $X$ takes
numeric values, the \emph{$k$th moment} of $d$ is defined as
\[
  G(d)_k \triangleq \Exp{ X^k }
\]
for $k \in \Nats$.  If $X$ ranges over tuples, the \emph{correlations} of $d$
are defined as
\[
  G(d, \{ a, b \})_{p, q}
  \triangleq
  \Exp{ \pi_a(X)^p \cdot \pi_b(X)^q } ,
\]
for $p, q \in \Nats$, and similarly for products of three or more projections.
Here, the \emph{projection} $\pi_i(X)$ for $X$ a tuple-valued random variable is
the marginal distribution of the $i$th coordinate of the tuple.

\paragraph*{Martingales.}

A martingale is a stochastic process with a special property: the average value
of the current step is equal to the value of the previous step.

\begin{defi}
  Let $\{ X_i \}$ be a stochastic process adapted to filtration $\{
  \mathcal{F}_i \}$. We say that $X$ is a \emph{martingale} with respect to
  $\mathcal{F}$ if it satisfies the property
  \[
    \CondExp{ X_i }{ \mathcal{F}_{i - 1} } = X_{ i - 1 } .
  \]
\end{defi}

For a simple example, consider a symmetric random walk on the
integers. Let $X \in \Ints$ denote the current position of the walk. At
each step, we flip a fair coin: if heads, we increase the position by $1$,
otherwise we decrease the position by $1$. The sequence of positions $X_0, X_1,
\dots$ forms a martingale since the average position at time $i$ is simply the
position at time $i - 1$:
\[
  \CondExp{ X_i }{ \mathcal{F}_{i - 1} }
  = X_{i - 1} .
\]
\paragraph*{Doob's decomposition.}

One important result in martingale theory is Doob's decomposition.
Informally, it establishes that any integrable random process can be
written uniquely as a sum of a martingale and a predictable process.
For our purposes, it gives a constructive and purely symbolic method to extract
a martingale from any arbitrary random process.

\begin{thm}[Doob's decomposition] \label{thm:doob}
  Let $X = \{ X_i \}_{i \in \Nats}$ be a stochastic process adapted to
  filtration $\{ \mathcal{F}_i \}_{i \in \Nats}$ where each $X_i$ has finite
  expected value. Then, the following process is a martingale:
  \[
    M_i =
    \begin{cases}
      X_0 &: i = 0 \\
      X_0 + \sum_{j = 1}^i X_j - \CondExp{ X_j }{ \mathcal{F}_{j - 1} } &: i > 0
    \end{cases}
  \]
  If $X$ is already a martingale, then $M = X$.
\end{thm}

We will think of the stochastic process $X$ as a seed process which generates
the martingale. While the definition of the martingale involves conditional
expectations, we will soon see how to automatically simplify these expectations.

\paragraph*{Optional stopping theorem.}

For any martingale $M$, it is not hard to show that the expected value of $M$
remains invariant at each time step. That is, for any fixed value $n \in \Nats$,
we have
\[
  \Exp{ M_n } = \Exp{ M_0 } .
\]
The optional stopping theorem extends this equality to situations where $n$
itself may be random, possibly even a function of the martingale.

\begin{defi}
  Let $(\Omega, \SigmaAlgebra, \Prob)$ be a probability space with filtration
  $\{ \mathcal{F}_i \}_{i \in \Nats}$. A random variable $\tau : \Omega \to
  \Nats$ is a \emph{stopping time} if the subset $\{ w \in \Omega \mid \tau(w)
  \leq i \}$ is a member of $\mathcal{F}_i$ for each $i \in \Nats$.
\end{defi}

Returning to our random walk example, the first time that the position is
farther than $100$ from the origin is a stopping time since this time depends
only on past samples. In contrast, the last time that a position is farther than
$100$ from the origin is \emph{not} a stopping time, since this time depends on
future samples.  More generally, the iteration count when we exit a
probabilistic loop is a stopping time since termination is a function of past
samples only.

If we have a stopping time and a few mild conditions, we can apply the optional
stopping theorem.\footnote{%
  A basic version of the optional stopping theorem will suffice for our purposes,
  but there are alternative versions that don't require finite expected stopping
  time and bounded increments.}

\begin{thm}[Optional stopping] \label{thm:OST}
  Let $\tau$ be a stopping time, and let $M$ be a martingale. If the expected
  value of $\tau$ is finite, and if $|M_i - M_{i-1}| \leq C$ for all $i > 0$ and
  some constant $C$, then
  \[
    \Exp{ M_\tau } = \Exp{ M_0 } .
  \]
\end{thm}

To see this theorem in action, consider the random walk martingale $S$ and
take the stopping time $\tau$ to be the first time that $|S| \geq 100$. It is
possible to show that $\tau$ has finite expected value, and clearly $|S_i - S_{i
  - 1}| \leq 1$. So, the optional stopping theorem gives
\[
    \Exp{ S_\tau } = \Exp{ S_0 } = 0 .
\]
Since we know that the position is $\pm 100$ at time $\tau$, this immediately
shows that the probability of hitting $+100$ is equal to the probability of
hitting $-100$. This intuitive fact can be awkward to prove using standard probabilistic
invariants, but falls out directly from a martingale analysis.

\section{Overview of method} \label{sec:tool}
Now that we have seen the key technical ingredients of our approach,
let us see how to combine these tools into an automated program
analysis.  We will take an imperative program specifying a stochastic
process and a seed expression, and we will automatically synthesize a
martingale and an assertion that holds at termination. We proceed in three
stages: extracting a polynomial representing the stochastic process in the
program, applying Doob's decomposition to the polynomial representation, and
applying optional stopping. We perform symbolic manipulations to simplify the
martingale and final fact.

\paragraph*{Programs.}
We consider programs of the form:\footnote{%
  We focus on programs for which our method achieves full automation. For
  instance, we exclude conditional statements because it is difficult to fully
  automate the resulting simplifications. We note however that there are
  standard transformations for eliminating conditionals; one such transformation
  is \emph{if-conversion}, a well-known compiler
  optimization~\citep{Allen:1983}.}
$$
  I; \WWhile{e}{(S; B)}
$$
where $I$ and $B$ are sequences of deterministic assignments (of the
form $\Ass{\Var}{\Expr}$), and $S$ is a sequence of probabilistic
samplings (of the form $\Rand{\SVar}{\DExpr}$).

Note that we separate \emph{sample variables} $s \in \SVar$, which are the
target of random samplings, from \emph{process variables} $x \in \Var$, which
are the target of deterministic assignments. This distinction will be important
for our simplifications: we know the moments and correlations of sample
variables, while we have less information for process variables. We require that
programs assign to sample variables before assigning to process variables in
each loop iteration; this restriction is essentially without loss of generality.

We take $\DExpr$ to
be a set of standard distributions over the integers or over finite
tuples of integers, to model joint distributions. 
For instance, we often consider the distribution $\mathsf{Bern}(1/2, \{-1, 1\})
\in \DExpr$ that returns $-1$ and $+1$ with equal probability.
We assume that all distributions in $\DExpr$ have bounded support; all
moments and correlations of the primitive distributions are finite.
We will also assume that distributions do not depend on the program state.

The set $\Expr$ of expressions is mostly standard, with a few notational
conveniences for defining stochastic processes:
\[
\begin{array}{r@{\ \ }l@{\quad}l}
\Expr ::= & \Var
     \mid  \SVar       
     \mid  \Var[-n]   & \mbox{process/sample/history variables} \\
     \mid& \Ints         & \mbox{constants}\\
     % \mid& \indFn{\Expr}        & \mbox{indicator}\\
     \mid& \pi_a(\Expr)        & \mbox{projections}\\
     \mid& \Expr + \Expr \mid \Expr \cdot \Expr        & \mbox{arithmetic}\\
     \mid& \Expr < \Expr \mid \Expr \land \Expr \mid \neg \Expr & \mbox{guards}\\
\end{array}
\]

\emph{History variables} $\Var[-n]$ are indexed by a positive integer
$n$ and are used inside loops. The variable $x[-n]$ refers to the
value of $x$ assigned $n$ iterations before the current iteration. If
the loop has run for fewer than $n$ iterations, $x[-n]$ is specified
by the initialization step of our programs:
\[
  I \triangleq
  \Ass{x_1[-n_1]}{e_1} ;
  \cdots ;
  \Ass{x_k[-n_k]}{e_k} .
\]

\paragraph*{Extracting the polynomial.}

For programs in our fragment, each variable assigned in the loop determines a
stochastic process: $x$ is the most recent value, $x[-1]$ is the previous value,
etc.  In the first stage of our analysis, we extract polynomial representations
of each stochastic process from the input program.

We focus on the variables that are mutated in $B$---each of these variables
determines a stochastic process. To keep the notation light, we will explain our
process for \emph{first-order} stochastic processes: we only use history
variables $x[-1]$ from the past iteration. We will also suppose that there is
just one process variable and one sample variable, and only samples from the
current iteration are used.

Since our expression language only has addition and multiplication as operators,
we can represent the program variable $x$ as a polynomial of other program
variables:
\begin{equation} \label{eq:p-poly}
  x = P_x(x[-1], s)
\end{equation}

Next, we pass to a symbolic representation in terms of (mathematical) random
variables. To variable $x$, we associate the random variable $\{ X_i \}_{i \in
  \Nats}$ modeling the corresponding stochastic process, and likewise for the
sample variable $s$. By convention, $i = 0$ corresponds to the initialization
step, and $i > 0$ corresponds to the stochastic process during the loop. In
other words,
\[
  \Ass{x[0]}{0}; \WWhile{e}{\Rand{s}{d}; \Ass{x}{x[-1] + s}}
\]
desugars to
\[
  \Ass{x[0]}{0}; \Ass{i}{0};
  \WWhile{e}{\Ass{i}{i + 1}; \Rand{s}{d}; \Ass{x[i]}{x[i-1] + s}}
\]
in a language with arrays instead of history variables. Then, the
program variable $x[i]$ corresponds to the random variable $X_i$.

Then, \Cref{eq:p-poly} and the initial conditions specified by the command $I$
give an inductive definition for the stochastic process:
\begin{equation} \label{eq:p-poly-rv}
  X_i = P_x(X_{i - 1}, S_i) .
\end{equation}

\paragraph*{Applying Doob's decomposition.}

The second stage of our analysis performs Doob's decomposition on the symbolic
representation of the process. We know that the seed expression $e$
must be a polynomial, so we can form the associated stochastic process $\{ E_i
\}_{i \in \Nats}$ by replacing program variables by their associated random
variable:
\begin{equation} \label{eq:p-poly-seed}
  E_i = P_e(X_i, S_i)
  .
\end{equation}
(Recall that the initial conditions $X_0$ and $S_0$, which define $E_0$, are
specified by the initialization portion $I$ of the program.)

Then, Doob's decomposition produces the martingale:
\[
  M_i = 
  \begin{cases}
    E_0 &: i = 0 \\
    E_0 + \sum_{j = 1}^i E_j - \CondExp{ E_j }{ \mathcal{F}_{j - 1} } &: i > 0 .
  \end{cases}
\]
To simplify the conditional expectation, we unfold $E_j$ via
\Cref{eq:p-poly-seed} and unroll the processes $X_i$ by one step with
\Cref{eq:p-poly-rv}.

\begin{figure}[t]
  \begin{align*}
    \CondExp{ c \cdot f + c' \cdot g }{ - }
    &\mapsto c \cdot \CondExp{ f }{ - } + c' \cdot \CondExp{ g }{ - }
  \end{align*}
  
  \hrule

  \begin{align}
  \CondExp{ X_{i - n} \cdot f }{ \mathcal{F}_{i - 1} }
  &\mapsto X_{i - n} \cdot \CondExp{ f }{ \mathcal{F}_{i - 1} }
  \tag{$n > 0$} \\
  \CondExp{ S_i \cdot S_i' }{ \mathcal{F}_{i - 1} }
  &\mapsto \CondExp{ S_i }{ \mathcal{F}_{i - 1} } \cdot \CondExp{ S_i' }{ \mathcal{F}_{i - 1} }
  \tag{$S \neq S'$}
  \end{align}

  \hrule

  \begin{align}
  \CondExp{ S^k_i }{ \mathcal{F}_{i - 1} }
  &\mapsto G(d)_k
  \tag{$S \sim d$} \\
  \CondExp{ \pi_a(S_i)^p \cdot \pi_b(S_i)^q }{ \mathcal{F}_{i - 1} }
  &\mapsto G(d_{a, b})_{p, q}
  \tag{$S \sim d$}
  \end{align}
  \caption{Selection of simplification rules}
  \label{fig:simex}
\end{figure}

Now, we apply our simplification rules; we present a selection in
\Cref{fig:simex}. The rules are divided into three groups (from top): linearity
of expectation, conditional independence, and distribution information.  The
first two groups reflect basic facts about expectations and conditional
expectations. The last group encodes the moments and correlations of the
primitive distributions. We can pre-compute these quantities for each primitive
distribution $d$ and store the results in a table.

By the form of \Cref{eq:p-poly-seed}, the simplification removes all
expectations and we can give an explicit definition for the martingale:
\begin{equation} \label{eq:q-poly}
  M_i =
  \begin{cases}
    E_0 &: i = 0 \\
    Q_e(X_{i - 1}, \dots, X_0, S_i, \dots, S_1) &: i > 0 ,
  \end{cases}
\end{equation}
where $Q_e$ is a polynomial.

\paragraph*{Applying optional stopping.}

With the martingale in hand, the last stage of our analysis applies
the optional stopping theorem. To meet the technical conditions of the
theorem, we need two properties of the loop:
\begin{itemize}
  \item The expected number of iterations must be finite.
  \item The martingale must have bounded increments.
\end{itemize}
These side conditions are non-probabilistic assertions that can
already be handled using existing techniques. For instance, the first
condition follows from the existence of a \emph{bounded
variant}~\citep{HartSP83}: an integer expression $v$ such that
\begin{itemize}
  \item $0 \leq v < K$; 
  \item $v = 0$ implies the guard is false; and
  \item the probability that $v$ decreases is strictly bigger than $\epsilon$
\end{itemize}
throughout the loop, for $\epsilon$ and $K$ positive constants. However in
general, finding a bounded variant may be difficult; proving finite expected
stopping time is an open area of research which we do not address here.

The second condition is also easy to check. For one possible approach, one can
replace stochastic sampling by non-deterministic choice over the support of
the distribution, and verify that the seed expression $e$ is bounded using
standard techniques~\citep{CousotH78,Mine06,MouraB08}. This suffices to show that the martingale
$M_i$ has bounded increments. To see why, suppose that the seed expression is
always bounded by a constant $C$. By Doob's decomposition, we have
\begin{align*}
  | M_i - M_{i - 1} |
  &= \left|\left(\sum_{j = 1}^i E_j - \CondExp{ E_j }{ \mathcal{F}_{j - 1} }\right)
  - \left(\sum_{j = 1}^{i - 1} E_j -
    \CondExp{ E_j }{ \mathcal{F}_{j - 1} }\right) \right| \\
  &= \left|E_i - \CondExp{ E_i }{ \mathcal{F}_{j - 1} }\right|
  \leq 2C ,
\end{align*}
so the martingale has bounded increments.

Thus, we can apply the optional stopping theorem to \Cref{eq:q-poly}
to conclude:
\[
  E_0
  = \Exp{M_0}
  =\Exp{M_\tau}
  = \Exp{Q_e(X_{\tau - 1}, \dots, X_0, S_\tau, \dots, S_1)} 
\]
Unlike the simplification step after applying Doob's decomposition, we
may not be able to eliminate all expected values. For instance, there
may be expected values of $X$ at times before $\tau$. However, if we
have additional invariants about the loop, we can often simplify the
fact with basic symbolic transformations.

\paragraph*{Implementation.}

\begin{figure}[t]
  \includegraphics[width=\textwidth]{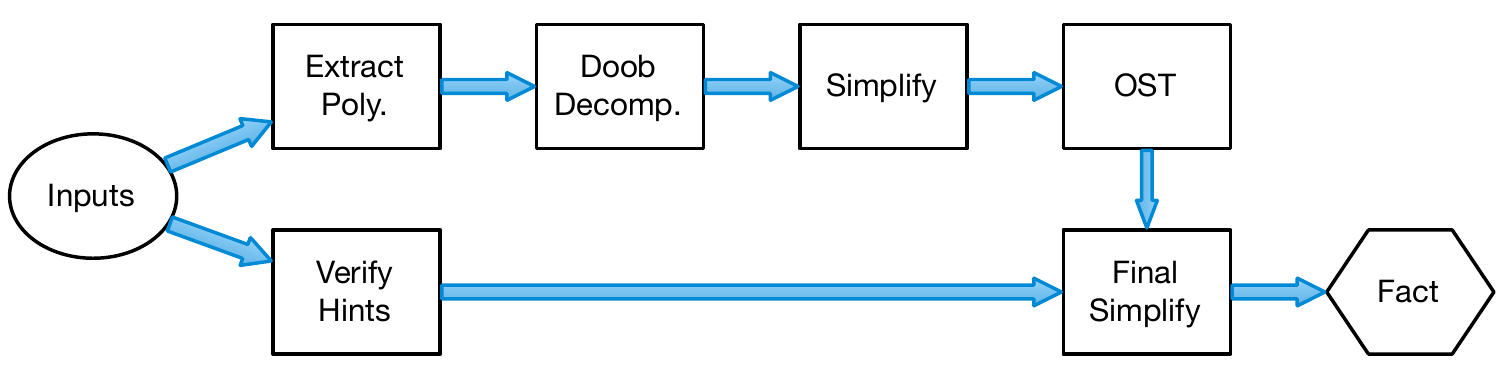}
  \caption{Tool pipeline}
  \label{fig:pipeline}
\end{figure}

We have implemented our process in a Python prototype using the
\texttt{sympy} library for handling polynomial and summation 
manipulations~\citep{Joyner:2012}. \Cref{fig:pipeline} shows the
entire pipeline.  There are three parts of the input: the program
describing a stochastic process, the seed expression, and hint
facts. The output is a probabilistic formula that holds at
termination.

The most challenging part of our analysis is the last stage: applying
OST.  First, we need to meet the side conditions of the optional
stopping theorem: finite expected iteration count and bounded
increments. Our prototype does not verify these side conditions
automatically since available termination tools are either not fully
automatic \citep{EsparzaGK12} or can only synthesize linear ranking
supermartingales \citep{ChakarovS13, ChatterjeeFNH16} that are insufficient
for the majority of our case studies\footnote{Although most of the ranking
supermatingales needed in our case studies are non-linear, the bounded
variants are always linear.}.
Furthermore, the final fact typically cannot be simplified without
some basic information about the program state at loop termination. We
include this information as \emph{hints}. Hints are first-order
formulae over the program variables and the loop counter (represented
by the special variable $t$), and are used as auxiliary facts during
the final simplification step. Hints can be verified using standard
program verification tools since they are non-probabilistic. In our
examples, we manually translate the hints and the program into the input
language of EasyCrypt \citep{BartheGHB11}, and perform the verifications there.
Automating the translation poses no technical difficulty and is left for future work.

We note that the performance of the tool is perfectly reasonable for the
examples considered in the next section. For instance, it handles the
``ABRACADABRA'' example in less than 2 seconds on a modern laptop.

\section{Examples}

Now, we demonstrate our approach on several classic examples of stochastic
processes. In each case, we describe the code, the seed expression, and any
hints needed for our tool to automatically derive the final simplified fact.

\paragraph*{Geometric distribution.}

As a first application, we consider a program that generates a draw from the
geometric distribution by running a sequence of coin flips.
\begin{lstlisting}
x[0] := 0; 
while (z $\neq$ 0) do 
    z ~~ Bern(p, {1, 0});
    x := x[-1] + z;
end
\end{lstlisting}

Here, $\lstt{Bern(p, \{1, 0\})}$ is the distribution that returns $1$ with
probability $p > 0$, and $0$ otherwise. The program simply keeps drawing until
we sample $0$, storing the number of times we sample $1$ in $x$.

We wish to apply our technique to the seed expression $x$. First, we can extract
the polynomial equation:
\[
  X_i = X_{i - 1} + Z_i .
\]
Applying Doob's decomposition, our tool constructs and reduces the
martingale:
\[
  M_i =
  \begin{cases}
    X_0 &: i = 0 \\
    X_i - p \cdot i &: i > 0 .
  \end{cases}
\]

To apply optional stopping, we first need to check that the stopping time $\tau$
is integrable. This follows by taking $z$ as a bounded variant---it remains in
$\{ 0, 1 \}$ and decreases with with probability $p > 0$. Also, the martingale
$M_i$ has bounded increments: $|M_i - M_{i - 1}|$ should be bounded by a
constant. But this is clear since we can use a loop invariant to show that $|X_i
- X_{i - 1}| \leq 1$, and the increment is
\[
  |M_i - M_{i - 1}| = |X_i - X_{i - 1} - p|
  \leq |X_i - X_{i - 1}| + p
  \leq 1 + p .
\]
So, optional stopping shows that
\[
  0 = \Exp{ X_\tau - p \cdot \tau } .
\]
With the hint $x = t - 1$---which holds at termination---our tool replaces
$X_\tau$ by $\tau - 1$ and automatically derives the expected running time:
\begin{align*}
  0 &= \Exp{ \tau - 1 - p \cdot \tau } \\
  \Exp{ \tau } &= 1/(1-p) .
\end{align*}

\paragraph*{Gambler's ruin.}

Our second example is the classic \emph{Gambler's ruin} process.  The program
models a game where a player starts with $a > 0$ dollars and keeps tossing a
fair coin. The player wins one dollar for each head and loses one dollar for
each tail. The game ends either when the player runs out of money, or reaches
his target of $b > a$ dollars. We can encode this process as follows:

\begin{lstlisting}
x[0] := a; 
while (0 < x < b) do 
    z ~~ Bern(1/2, {-1, 1});
    x := x + z;
end
\end{lstlisting}

We will synthesize two different martingales from this program, which will yield
complementary information once we apply optional stopping. For our first
martingale, we use $x$ as the seed expression. Our tool synthesizes the
martingale
\[
  M_i =
  \begin{cases}
    X_0 &: i = 0 \\
    X_i &: i > 0 .
  \end{cases}
\]
So in fact, $x$ is already a martingale.

To apply optional stopping, we first note that $x$ is a bounded variant: it
remains in $(0, b)$ and decreases with probability $1/2$ at each iteration.
Since the seed expression $x$ is bounded, the
martingale $M_i$ has bounded increments. Thus, optional stopping yields
\[
  a = \Exp{ X_0 } = \Exp{ X_\tau } .
\]
If we give the hint
\begin{equation} \label{eq:ruin-inv}
  (x = 0) \lor (x = b)
\end{equation}
at termination, our prototype automatically derives
\begin{align*}
  a &= \Exp{ X_\tau \cdot \indFn{X_\tau = 0 \lor X_\tau = b} } \\
  &= \Exp{ X_\tau \cdot \indFn{ X_\tau = 0 } } + \Exp{ X_\tau \cdot \indFn{ X_\tau = b } } \\
  &= 0 \cdot \Pr{}{ X_\tau = 0 } + b \cdot \Pr{}{ X_\tau = b }
  = b \cdot \Pr{}{ X_\tau = b } ,
\end{align*}
so the probability of exiting at $b$ is exactly $a/b$.

Now, let us take a look at a different martingale generated by the seed
expression $x^2$. Our prototype synthesizes the following martingale:
\[
  M_i' =
  \begin{cases}
    X_0^2 &: i = 0 \\
    X_i^2 - i &: i > 0
  \end{cases}
\]
Again, we can apply optional stopping: $x$ is a bounded variant, and the seed
expression $x^2$ remains bounded in $(0, b^2)$. So, we get
\[
  a^2 = \Exp{ M_0 } = \Exp{ X_\tau^2 - \tau } .
\]
By using the same hint \Cref{eq:ruin-inv}, our prototype automatically derives
\begin{align*}
  a^2 &= \Exp{ X_\tau^2 \cdot \indFn{X_\tau = 0 \lor X_\tau = b} } - \Exp{ \tau } \\
  &= \Exp{ X_\tau^2 \cdot \indFn{ X_\tau = 0 } } + \Exp{ X_\tau^2 \cdot \indFn{
      X_\tau = b } } - \Exp{ \tau } \\
  &= 0 \cdot \Pr{}{ X_\tau = 0 } + b^2 \cdot \Pr{}{ X_\tau = b } - \Exp{ \tau }
  = b^2 \cdot \Pr{}{ X_\tau = b } - \Exp{ \tau } .
\end{align*}
Since we already know that $\Pr{}{ X_\tau = b } = a/b$ from the first martingale
$\{ M_i \}_{i \in \Nats}$, this implies that the expected running time of the
Gambler's ruin process is
\[
  \Exp{ \tau } = a(b - a) .
\]

\paragraph*{Gambler's ruin with momentum.}
Our techniques extend naturally to stochastic processes that depend on variables
beyond the previous iteration. To demonstrate, we'll consider a variant of
Gambler's ruin process with momentum: besides just the coin flip, the gambler
will also gain profit equal to the difference between the \emph{previous two}
dollar amounts. Concretely, we consider the following process:
\begin{lstlisting}
x[0] := a;
x[1] := a;
while (0 < x < b) do 
  z ~~ Bern(p, {-1, 1});
  x := x[-1] + (x[-1] - x[-2]) + z;
end
\end{lstlisting}
Note that we must now provide the initial conditions for two steps, since the
process is second-order recurrent. Given seed expression $x$, our tool
synthesizes the following martingale:
\[
  M_i =
  \begin{cases}
    X_0 &: i = 0 \\
    X_0 + X_i - X_{i - 1} &: i > 0
  \end{cases}
\]
Identical to the Gambler's ruin process, we can verify the side conditions and
apply optional stopping, yielding
\[
  a = \Exp{ M_0 } = \Exp{ M_\tau } = \Exp{ X_0 + X_\tau - X_{\tau - 1} } .
\]
Unfolding $X_0 = a$ and simplifying, our tool derives the fact
\[
  \Exp{ X_\tau } = \Exp{ X_{\tau - 1} } .
\]
We are not aware of existing techniques that can prove this kind of
fact---reasoning about the expected value of a variable in the iteration
\emph{just prior} to termination.

\paragraph*{Abracadabra.}

Our final example is a classic application of martingale reasoning. In this
process, a monkey randomly selects a character at each time step, stopping when
he has typed a special string, say ``ABRACADABRA''. We model this process as
follows:
\begin{lstlisting}
match$_0$[0] :=  1;
match$_1$[0] :=  0;
...
match$_{11}$[0] :=   0;
while (match$_{11}$ == 0) do
    s ~~ UnifMatches;
    match$_{11}$ :=   match$_{10}$[-1] * $\pi_{11}$(s);
    match$_{10}$ :=   match$_{9}$[-1] * $\pi_{10}$(s);
    ...
    match$_{1}$ :=   match$_{0}$[-1] * $\pi_1$(s);
end
\end{lstlisting}
Here, $\lstt{UnifMatches}$ is a distribution over tuples that represents a
uniform $c$ draw from the letters, where the $k$th entry is $1$ if the $c$
matches the $k$th word and $0$ if not. The variables $\lstt{match}_i$ record
whether the $i$ most recent letters match the first $i$ letters of target word;
$\lstt{match}_0$ is always $1$, since we always match $0$ letters.

Now, we will apply Doob's decomposition. Letting $L$ be the number of possible
letters and taking the seed expression
\[
  e \triangleq 1 + L \cdot \lstt{match}_{1}
  + \cdots
  + L^{11} \cdot \lstt{match}_{11} ,
\]
our tool synthesizes the following martingale:
\[
  M_i =
  \begin{cases}
    1 + L \cdot X^{(1)}_i + \cdots + L^{11} \cdot X^{(11)}_i &: i = 0 \\
    \sum_{j = 1}^i \left(
      L \cdot X^{(1)}_j + \cdots + L^{11} \cdot X^{(11)}_j
      -
      L^0 \cdot X^{(0)}_{j - 1} + \cdots + L^{10} \cdot X^{(10)}_{j - 1}
    \right)
    &: i > 0 ,
  \end{cases}
\]
where $X^{(j)}$ is the stochastic process corresponding to $\lstt{match}_j$. The
dependence on $L$ is from the expectations of projections of \lstt{UnifMatches},
which are each $1/L$---the probability of a uniformly random letter matching any
fixed letter.

To apply the optional stopping theorem, note that the seed expression $e$ is
bounded in $(0, L^{12})$, and $1 + L + \cdots + L^{11} - e$ serves as a bounded
variant: take the highest index $j$ such that $\lstt{match}_j = 1$, and there is
probability $1/L$ that we increase the match to get $\lstt{match}_{j + 1} = 1$,
decreasing the variant. So, we have
\[
  1 = \Exp{ M_0 } =
    \sum_{j = 1}^\tau
    \Exp{ L \cdot X^{(1)}_j + \cdots + L^{11} \cdot X^{(11)}_j }
    -
    \Exp{ L^0 \cdot X^{(0)}_{j - 1} + \cdots + L^{10} \cdot X^{(10)}_{j - 1} } .
\]
Our tool simplifies and uses the hints $X^{(11)}_j = 0$ and $X^{(0)}_j = 1$ for
$j < \tau$ to derive
\[
  1 =  L^0 \cdot \Exp{ X^{(0)}_{\tau} }
  + \cdots +
  L^{11} \cdot \Exp{ X^{(11)}_{\tau} } - \Exp{ \tau } .
\]
For the target string ``ABRACADABRA'', we use hints
\begin{align}
  (\lstt{match}_{11} = 1) &\implies (\lstt{match}_{4} = 1)
  \notag \\
  (\lstt{match}_{11} = 1) &\implies (\lstt{match}_{1} = 1)
  \notag \\
  (\lstt{match}_{11} = 1) &\implies (\lstt{match}_{0} = 1)
  \notag \\
  (\lstt{match}_{11} = 1) &\implies (\lstt{match}_{j} = 0)
  \tag{for $j \neq 0, 1, 4, 11$} .
\end{align}
For example, if $\lstt{match}_{11}$ is set then the full string is matching
``ABRACADABRA'', so the most recently seen four characters are ``ABRA''. This
matches the first four letters of ``ABRACADABRA'', so $\lstt{match}_4$ is also
set. The hint can be proved from a standard loop invariant.

Our tool derives the expected running time:
\[
  \Exp{ \tau } = L^{1} + L^{4} + L^{11} .
\]
\subsection*{Benchmarks.}

To give an idea of the efficiency of our tool, we present some benchmarks for
our examples in \Cref{fig:bench}. We measured timing on a recent laptop with a
2.4 GHz Intel Core processor with 8 GB of RAM. We did not optimize for the
performance; we expect that the running time could be greatly improved with some
tuning.

\begin{table}[h]
  \centering
\begin{tabular}{|c|c|}
  \hline
  Example           & Running time (s) \\
  \hline
  \textsc{geom}     & 0.14             \\
  \textsc{gamble}   & 0.11             \\
  \textsc{gamble2}  & 0.17             \\
  \textsc{miniabra} & 0.87             \\
  \textsc{fullabra} & 3.58             \\
  \hline
\end{tabular}
\caption{Preliminary benchmarks.}
\label{fig:bench}
\end{table}
The example \textsc{miniabra} is a smaller version of the ABRACADABRA example,
where the alphabet is just $\{ 0, 1 \}$, and we stop when we sample the sequence
$111$; \textsc{fullabra} is the full ABRACADABRA example.

While there is a growing body of work related to martingale techniques for
program analysis (see the following section), it is not obvious how to compare
benchmarks. Existing work focuses on searching for martingale expressions within
some search space; this is a rather different challenge than synthesizing a
single martingale from a programmer-provided seed expression. In particular, if
the seed expression happens to already be a martingale by some lucky guess, our
tool will simply return the seed expression after checking that it is indeed a
martingale.

\section{Related work}\label{sec:related}

\paragraph*{Martingales.}
Martingale theory is a central tool in fields like statistics, applied
mathematics, control theory, and finance.  When analyzing
randomized algorithms, martingales can show tight bounds on tail events
\citep{MotwaniR95}.  In the verification community, martingales are used as
invariant arguments, and as variants arguments to prove almost sure termination
\citep{BournezG05,ChakarovS13,FerrerH15,ChatterjeeFNH16}.
Recently, martingale approaches were extended to prove more complex
properties.
\citet{ChakarovVS16} propose proof rules for proving persistence and recurrence
properties. \citet{DimitrovaFHM16} develop a deductive proof system for
PCTL$^*$, with proof rules based on martingales and supermartingales.

\paragraph*{Probabilistic Hoare logic.}
\citet{McIverMorgan05} propose a Hoare-like logic that is
quite similar to our approach of using martingales and OST. Their approach is
based on
\emph{weakest pre-expectations}, which are an extension of Dijkstra's weakest
preconditions \citep{Dijkstra75} based on ``backward'' conditional expectations.
Their probabilistic invariants are similar to submartingales, as the expected
value of the invariant at the beginning of the execution lower bounds the expected
value of the invariant at termination. Their proof rule also requires an
additional constraint to ensure soundness, but it
requires a limiting argument that is more difficult to automate compared to
our bounded increment condition. We could relax our condition using a weaker
version of OST that generalizes their condition \citep{Williams91}.
Another substantial difference with our approach is that their logic supports
non-deterministic choices---ours does not. It is not obvious how we can extend
our synthesis approach to the non-probabilistic case as we heavily rely on the
concept of filtration, not applicable in the presence of non-determinism.

\paragraph*{Probabilistic model checking.}
In the last twenty years, model checking technology for probabilistic models
have made impressive strides \citep{KwiatkowskaNP11,CiesinskiB06,KatoenZHHJ11}
(\citet{BaierKatoenBook} provide overview).  The main advantage of model
checking is that it requires nothing from the user;
our technique requires a seed expression.
However, model checking techniques suffer from the state explosion problem---the
time and memory consumption of the model checking algorithm depends on the
number of reachable states of the program.
Our approach can be used to verify infinite and parametric programs without
any performance penalty, as we work purely symbolically.
For example, a probabilistic model checker can find the expected running time of
the gambler's ruin process for concrete values of \(a\) and \(b\) but they
cannot deduce the solution for the general case, unlike our technique.

\paragraph*{Invariant synthesis.}
There are several approaches for synthesizing probabilistic invariants.
\citet{KatoenMMM10} propose the first complete method for the
synthesis of McIver and Morgan's probabilistic linear invariants. It is an
extension of the constraint solving approach by \citet{ColonSS03}
for the synthesis of (non-probabilistic) linear invariants.
\citet{ChakarovS13} later extended this work
to martingales and ranking supermartingales.
\citet{ChakarovS14} propose a new notion of
probabilistic invariants that generalizes the notion of supermatingales. They
give a synthesis approach based on abstract interpretation, but it is not
clear how their techniques can prove properties at termination time.
\citet{ChenHWZ15} propose a synthesis tool for verifying
Hoare triples in the McIver and Morgan logic, using a combination of
Lagrange's interpolation, sampling, and counterexample guided search.  One of
the novelties is that they can synthesize non-linear invariants.  The main
disadvantages is that one must manually check the soundness condition, and one
must provide a pre-expectation.
For instance, we can apply the method of \citet{ChenHWZ15} to the gambler's ruin
process only if we already know that the expected running time is \(a(b - a)\).
In contrast, we can deduce \(\Exp{\tau} = a(b-a)\) knowing only that
\(\Exp{\tau}\) is finite.

\paragraph*{Expected running time.}
As the termination time of a probabilistic program is a random quantity, it is natural
to measure its performance using the average running time. Rough bounds can be obtained
from martingale-based termination proofs \citep{FerrerH15}.
Recently, \citet{ChatterjeeFNH16} showed that arbitrary approximations can be obtained
from such proofs when the program is linear. They use Azuma's inequality to obtain a tail
distribution of the running time, and later they model check a finite unrolling of the loop.
\citet{Monniaux01} propose a similar approach that uses abstract interpretation to obtain
the tail distribution of the running time.
\citet{KaminskiKMO16} extend Nielson's proof system \citep{Nielson87} to
bound the expected running time of probabilistic programs.

\paragraph*{Recurrence analysis.}
Our synthesis approach is similar to the use of recurrences relations for the
synthesis of non-probabilistic invariants
\citep{AmmarguellatH90,RodriguezK07,Kovacs08}.  The main idea is to find
syntactic or semantic recurrences relations, and later simplify them using known
closed forms to obtain loop invariants.  In essence, we apply algebraic
identities to simplify the complex martingales from Doob's decomposition.  The
difference is that our simplifications are more complex as we cannot always
obtain a closed form but a simpler summation.  However, we obtain the same
closed form when we apply Doob's decomposition to inductive variables.
Another difference is that we rely on the syntactic criteria to
identify which values are predictable and which values are random.

\section{Conclusion}
We proposed a novel method for automatically synthesizing martingales
expressions for stochastic processes.  The basic idea is to transform
any initial expression supplied by the user into a martingale using
Doob's decomposition theorem.  Our method complements the
state-of-the-art synthesis approaches based on constraint solving.  On
one hand, we always output a martingale expression, we are able to
synthesize non-inductive martingales, and since we do not rely on
quantifier elimination, we can synthesize polynomial expression of
very high degree.  On the other hand, we do not provide any
completeness result, and the shape of martingale is difficult to
predict.

We considered several classical case studies from the literature,
combining our synthesis method with the optional stopping theorem and
non-probabilistic invariants to infer properties at termination time
in a fully automatic fashion. 

Future work includes extending our approach to programs with arrays and
improving the tool with automated procedures for checking side-conditions. It
would also be interesting to consider richer programs, say distributions with
parameters that depend on program state. Another possible direction would be
improving the simplification procedures; possibly, the tool could produce
simpler facts.  Experimenting with more advanced computer-algebra systems and
designing simplification heuristics specialized to handling the conditional
expectations synthesized by Doob's decomposition are both promising future
directions.  It would also be interesting to integrate our method as a special
tool in systems for interactive reasoning about probabilistic computations.

\paragraph*{Acknowledgments.}
We thank the anonymous reviewers for their helpful comments. This work was
partially supported by NSF grants TWC-1513694 and CNS-1065060, and by a grant
from the Simons Foundation ($\#360368$ to Justin Hsu).

\bibliographystyle{abbrvnat}
\bibliography{header,main,literature}

\end{document}

%% file: prelude.tex
\usepackage[english]{babel}
\usepackage[T1]{fontenc}
\usepackage[utf8]{inputenc} 
\usepackage{upgreek}
\usepackage{paralist}
\usepackage{enumitem}
\usepackage{caption}
\usepackage{mdframed}
\usepackage{subcaption}
\usepackage[numbers,longnamesfirst]{natbib}

\usepackage{xspace}
\usepackage{xcolor}
\usepackage[ruled,vlined]{algorithm2e}
\usepackage{nicefrac}
\usepackage{bbm}
\usepackage{graphicx}
\usepackage{mathpartir}
\usepackage{bbold}
\usepackage{yfonts}
\usepackage{amsmath,amsfonts,amssymb}

\usepackage{todonotes}

\usepackage{xargs}
\newtheorem{thm}{Theorem}
\newtheorem{defi}{Definition}

\newcommand{\Nats}{\mathbb{N}}
\newcommand{\Ints}{\mathbb{Z}}
\newcommand{\Reals}{\mathbb{R}}

\newcommand{\dif} {\mathrm{d}}
\newcommand{\Prob}{\mathbb{P}}

\newcommand{\ExpSymb}{\mathbf{E}}
\newcommand{\Exp}[1]{\ExpSymb[#1]}
\newcommand{\CondExp}[2]{\ExpSymb[#1 \mid #2]}
\newcommand{\indFn}[1]{\mathbb{1}_{\{#1\}}}
\newcommand{\SigmaAlgebra}[1][F]{\mathcal{#1}}

\newcommand{\Var}{\mathcal{X}}
\newcommand{\SVar}{\mathcal{S}}
\newcommand{\Expr}{\mathcal{E}}
\newcommand{\DExpr}{\mathcal{DE}}

\newcommand{\Ass}[2]{#1 \leftarrow #2}
\newcommand{\Rand}[2]{#1 \stackrel{\raisebox{-.25ex}[.25ex]%
 {\tiny $\mathdollar$}}{\raisebox{-.2ex}[.2ex]{$\leftarrow$}} #2}

\newcommand{\WWhile}[2]{\mathsf{while}\ #1\ \mathsf{do}\ #2}

\renewcommand{\Pr}[2]{\mathrm{Pr}_{#1}{\left[ #2 \right]}}

\usepackage{listings}

\usepackage[T1]{fontenc}
\usepackage{microtype}

\usepackage[scaled]{beramono}
\newcommand\Small{\fontsize{8.2pt}{8.4pt}\selectfont}
\newcommand*\LSTfont{\Small\ttfamily\SetTracking{encoding=*}{-60}\lsstyle}
\def\lstrnd{\stackrel{\raisebox{-.15ex}{\ensuremath{\scriptscriptstyle\$}}}{\raisebox{-.2ex}{\ensuremath{\leftarrow}}}}

\newcommand{\lstt}[1]{\mbox{\LSTfont #1}}

\usepackage{xcolor}
\definecolor{DarkGreen}{rgb}{0.1,0.5,0.1}
\definecolor{DarkRed}{rgb}{0.5,0.1,0.1}
\definecolor{DarkBlue}{rgb}{0.1,0.1,0.5}
\usepackage{hyperref}
\hypersetup{
    unicode=false,          % non-Latin characters in Acrobat's bookmarks
    pdftoolbar=true,        % show Acrobat toolbar?
    pdfmenubar=true,        % show Acrobat menu?
    pdffitwindow=false,      % page fit to window when opened
    pdftitle={},    % title
    pdfauthor={}
    pdfsubject={},   % subject of the document
    pdfnewwindow=true,      % links in new window
    pdfkeywords={keywords}, % list of keywords
    colorlinks=true,       % false: boxed links; true: colored links
    linkcolor=DarkRed,          % color of internal links
    citecolor=DarkGreen,        % color of links to bibliography
    filecolor=DarkRed,      % color of file links
    urlcolor=DarkBlue,          % color of external links
}

\usepackage{cleveref}
\crefname{section}{\S}{\S}
\Crefname{section}{\S}{\S}

\crefname{prop}{proposition}{propositions}
\Crefname{prop}{Proposition}{Propositions}

\crefname{lem}{lemma}{lemmas}
\Crefname{lem}{Lemma}{Lemmas}

\crefname{thm}{theorem}{theorems}
\Crefname{thm}{Theorem}{Theorems}

\crefname{defi}{definition}{definitions}
\Crefname{defi}{Definition}{Definitions}